\documentclass[
 reprint,
 superscriptaddress,
 amsmath,amssymb,
 aps
]{revtex4-2}

\usepackage[nopar]{lipsum}
\usepackage[dvipdfmx]{graphicx}
\usepackage{dcolumn}
\usepackage{bm}
\usepackage{color}

\usepackage{amsmath}

\begin{document}

\preprint{APS/123-QED}

\title{Negative correlation between the linear and the nonlinear conductance in magnetic tunnel junctions}

\author{Shuichi Iwakiri}
\email{shuichi@meso.phys.sci.osaka-u.ac.jp}
\affiliation{Department of Physics, Osaka University,1-1 Machikaneyamacho, Osaka, Japan}
\author{Satoshi Sugimoto}
\affiliation{Research Center for Magnetic and Spintronic Materials, National Institute for Materials Science (NIMS), 1-2-1 Sengen, Tsukuba, Japan}
\author{Yasuhiro Niimi}
\affiliation{Department of Physics, Osaka University,1-1 Machikaneyamacho, Osaka, Japan}
\affiliation{Center for Spintronics Research Network (CSRN), Osaka University, 1-3 Machikaneyamacho, Osaka, Japan}
\author{Yusuke Kozuka}
\affiliation{Research Center for Magnetic and Spintronic Materials, National Institute for Materials Science (NIMS), 1-2-1 Sengen, Tsukuba, Japan}
\author{Yukiko K. Takahashi}
\affiliation{Research Center for Magnetic and Spintronic Materials, National Institute for Materials Science (NIMS), 1-2-1 Sengen, Tsukuba, Japan}
\author{Shinya Kasai}
\affiliation{Research Center for Magnetic and Spintronic Materials, National Institute for Materials Science (NIMS), 1-2-1 Sengen, Tsukuba, Japan}
\affiliation{JST, PRESTO, 4-1-8 Honcho, Kawaguchi, Saitama, Japan}
\author{Kensuke Kobayashi}
\affiliation{Department of Physics, Osaka University,1-1 Machikaneyamacho, Osaka, Japan}
\affiliation{Institute for Physics of Intelligence and Department of Physics, The University of Tokyo, Bunkyo-ku, Tokyo, 113-0033, Japan}
\affiliation{Trans-scale Quantum Science Institute, The University of Tokyo, Bunkyo-ku, Tokyo, 113-0033 Japan}

\date{\today}

\begin{abstract}
The current-voltage ($IV$) characteristics beyond the linear response regime of magnetic tunnel junction (MTJ) is systematically investigated. We find a clear negative correlation between the two coefficients to characterize the linear ($I$$\propto$$V$) and the lowest-order nonlinear ($I$$\propto$$V^3$) currents, which holds regardless of the temperature and the thickness of the tunnel barrier. This observation cannot simply be explained by the standard tunneling model such as the Brinkman model, suggesting a mechanism intrinsic to MTJ. We propose a phenomenological model based on the Julliére model that attributes the observed negative correlation to the spin-flip tunneling assisted by a magnon. These results suggest a fundamental law between the linear and the nonlinear response of MTJ.
\end{abstract}

\maketitle
\section{\label{sec:level1}Introduction}
Magnetic tunnel junction (MTJ), consisting of ferromagnet-insulator-ferromagnet heterostructure plays a central role in spintronics.
Its resistance changes with the magnetization configuration from parallel (P) to anti-parallel (AP) known as the tunneling magnetoresistance (TMR) effect \cite{Julliere1975,Moodera1995,MIYAZAKI1995}.
In particular, the MTJ with MgO barrier exhibits a giant TMR of a few hundred \% \cite{Yuasa2004,Parkin2004,Djayaprawira2005}. Such giant TMR is due to the spin-dependence of both the density of states (DOS) in the electrode metal described by the Julliére model \cite{Julliere1975,PhysRevB.39.6995} and the decay rates depending on the Bloch states in the barrier \cite{Butler2001,PhysRevB.63.220403}.
However, it is commonly observed that the TMR is significantly suppressed by applying the bias voltage, as initially reported by Julliére \cite{Julliere1975}.
From the physical point of view, this effect stems from the bias voltage dependence of the conductance (nonlinear conductance) \cite{Tsymbal2003,Yuasa2007}, which is the main topic of this Article.

One of the known mechanisms to account for the nonlinear conductance in tunnel junctions is the modulation of the tunneling barrier height by the electric field. This can be addressed through models such as the Simmons model \cite{Simmons1963} or the Brinkman model \cite{Brinkman}, both of which are based on the WKB approximation of the tunneling. The Brinkman model is an extension of the Simmons model to include the asymmetry of the barrier height due to the difference of the electrode materials. It also gives a simple formula of the bias-dependent conductance at low bias regime.
This approach was applied to the nonlinear conductance in the P configuration of MTJ and the barrier height was estimated \cite{Moodera1998,Parkin2004}.

Other mechanisms to be responsible for the nonlinearity peculiar to MTJ are the electron's inelastic tunneling processes due to the interaction with quasiparticles (e.g. magnons and phonons) \cite{Zhang1997,Guinea1998,Bratkovsky,Moodera1998,LU2003205,Drewello2008,Drewello2009} or impurities \cite{Drewello2009,Appelbaum1967,Wei2010}.
Among them, tunneling with emitting/absorbing a magnon (magnon-assisted tunneling) has been extensively studied. For example, Zhang $et$ $al.$ proposed an analytical model of the magnon-assisted tunneling and obtained an agreement with the experiment in MTJ \cite{Zhang1997}. In addition, Moodera $et$ $al.$ found a peak/dip structure in the nonlinear conductance which was also attributed to the magnon \cite{Moodera1998}.

Those prior works have clarified the overall characteristics of the nonlinear conductance in MTJ. Especially, a lot is known about the nonlinear conductance at wide voltage range (up to a few hundreds of millivolts or tens of $k_{\rm B\it}T$ where $k_{\rm B\it}$ is Boltzmann constant and $T$ is temperature) and its behavior in either P or AP configuration.
On the other hand, the nonlinearity at lower bias regime (comparable to $k_{\rm B\it}T$) is relatively less understood.
For example, little has been known about the relation between the nonlinear conductance and the linear conductance and/or the magnetization configuration dependence (e.g., not only the P and AP states but also the intermediate ones).
Such understandings would enable us to understand the MTJ in more depth.

In this Article, the nonlinear electron transport in MTJ is systematically studied at low bias regime, controlling the magnetization configuration, the temperature, and the tunneling probability. We find that the two coefficients to characterize the linear ($I$$\propto$$V$) and the lowest-order nonlinear ($I$$\propto$$V^3$) currents are negatively correlated, which cannot be explained by the above barrier modulation effect described by the Brinkman model. Instead, we propose a phenomenological model based on the Julliére model, additionally taking the magnon-assisted tunneling into account. Our finding would extend the understanding of the nonlinear conductance in MTJ, which leads to the observed nontrivial negative correlations.
\section{\label{sec:level1}Experimental setup}
\begin{figure}[h]
\centering
\includegraphics[pagebox=cropbox,clip]{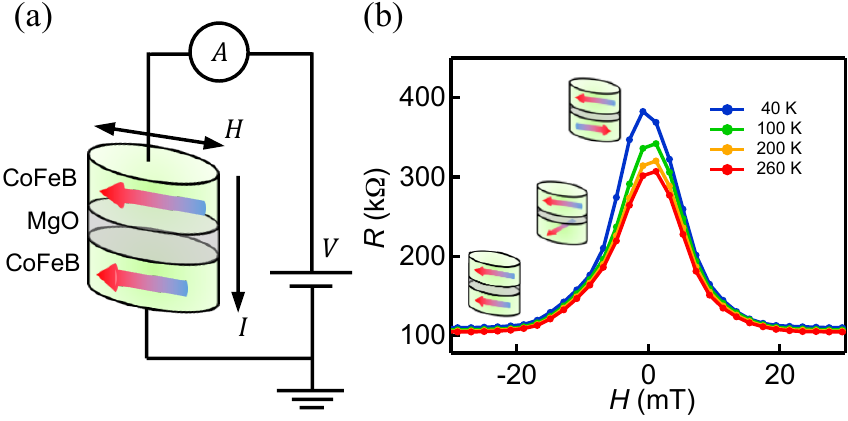}
\caption{(a) Schematic of the MTJ nanopillar. The magnetic field is applied along the long axis of the ellipse. (b) Magnetoresistance of the MTJ ($V=$ 5 mV, $d_{\rm MgO}=$ 1.33 nm) at temperature $T=$ 40, 100, 200, and 260 K. Only the sweep from positive to negative magnetic field is shown for simplicity. The schematic diagrams illustrate the magnetization configuration at each magnetic field.}
\label{pillar}
\end{figure}

Figure \ref{pillar}(a) shows the schematic of our sample. The multilayered stack consisting of Ta (5 nm) / Ru (10 nm) / Ta (5 nm) / Co$_{20}$Fe$_{60}$B$_{20}$ (5 nm) / MgO ($d_{\rm MgO}$ = 1.06, 1.13, 1.27, or 1.33 nm) / Co$_{20}$Fe$_{60}$B$_{20}$ (4 nm) / Ta (5 nm) / Ru (5 nm) is deposited on a thermally oxidized Si substrate by magnetron sputtering. The number in the parentheses is the thickness.

The film stack is annealed at 400 \textdegree {}C for 30 min. Hereafter, the ferromagnetic metal, Co$_{20}$Fe$_{60}$B$_{20}$, is denoted as CoFeB. The CoFeB / MgO / CoFeB layers form an MTJ with in-plane magnetic anisotropy as shown Fig. \ref{pillar}(a). The sample is patterned into an elliptical pillar of 150 nm×450 nm and the Ta (10 nm) / Au (100 nm) electrode is sputtered.
The thickness of the tunnel barrier $d_{\rm MgO}$ is set right above the minimal one to obtain the crystallinity at MgO / CoFeB interface \cite{Yuasa2004,Parkin2004}. This enables us to observe nonlinear transport at bias voltage as low as possible.
The $IV$ characteristics are obtained by applying voltage $V$, measuring current $I$ under in-plane magnetic field $H$ along the long axis of the pillar.

Figure \ref{pillar}(b) shows the magnetic field dependence of the MTJ resistance. Here, the data of $d_{\rm MgO}=$ 1.33 nm
and $T=$ 40, 100, 200, and 260 K are shown as a typical example. The bias voltage is $V=$ 5 mV, where the current is perfectly proportional to the voltage, and one can deduce the linear resistance ($R=V/I$). As the magnetic field is swept from positive to negative, the resistance takes a peak at $\sim$ 0 mT and it monotonically decreases as the field is increased. It finally saturates at around $\pm$20 mT.
According to the well-established phenomenological model (Julliére model) that connects the magnetoresistance to the spin-dependent density of states (DOS)
\cite{Julliere1975}, the resistance minimum $R_{\rm min}$ (maximum $R_{\rm max}$) corresponds to the magnetization configuration of P (AP) as illustrated in Fig. \ref{pillar}(b).
The magnetoresistance ratio $(R_{\rm max}-R_{\rm min})/R_{\rm min}$ is approximately 300 \%, which is a reasonable value for MgO-based MTJ \cite{Yuasa2004,Parkin2004}.
Unlike the steep switching behavior often realized by designing the MTJ structure, we intentionally design the present MTJs without inserting an antiferromagnetic layer which fixes the magnetization direction \cite{Yuasa2004,Parkin2004} so that their resistance gradually change as a function of the magnetic field.
This guarantees that the magnetization configuration is continuously controlled from P to AP in our samples.
Thus we can systematically investigate the relation between the nonlinear conductance and the magnetization configuration.

\section{\label{sec:level1}Experimental results}
\subsection{\label{sec:level2}Current-voltage characteristics}
\begin{figure}[h]
\begin{center}
\includegraphics[scale=1,pagebox=cropbox,clip]{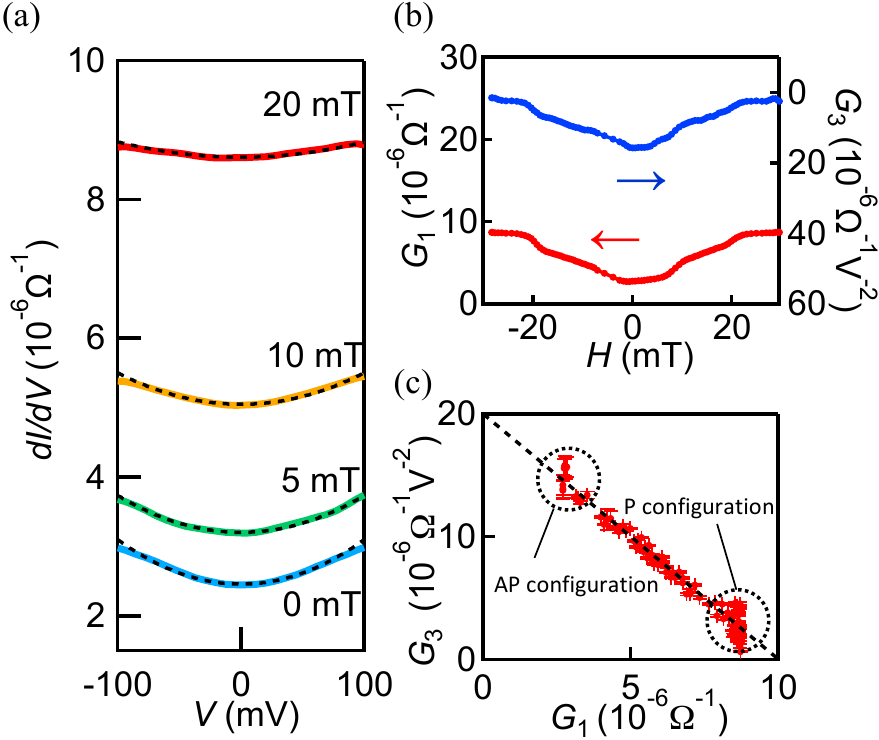}
\caption{Characterization of the nonlinear transport obtained for the MTJ with $d_{\rm MgO}=$ 1.33 nm at 260 K. (a) Bias dependence of the differential conductance $dI/dV$ at 0, 5, 10, and 20 mT. The black dashed curves are the result of the parabolic fitting. (b) Magnetic field dependence of the linear ($G_1$) and the nonlinear ($G_3$) conductance obtained from the fitting. (c) Correlation between $G_1$ and $G_3$. The areas of P and AP configurations are marked as black dotted circles.}
\label{fig2}
\end{center}
\end{figure}

The $IV$ characteristics for the samples with four different $d_{\rm MgO}$'s are measured ($d_{\rm MgO}$ = 1.06, 1.13, 1.27, and 1.33 nm), sweeping $V=\pm$ 100 mV, $H=\pm$30 mT, and $T=$40, 100, 200, and 260 K. By numerically differentiating the $IV$ characteristics, we obtain the differential conductance as shown in Fig. \ref{fig2}(a).
We show the data at 260 K and $d_{\rm MgO}$ =1.33 nm as an example.
As is clear from Fig. \ref{fig2}(a), the conductance is symmetric with respect to the sign reversal of a bias voltage up to $|V|=$ 100 mV.
%

The nonlinear component is well fitted by the parabolic function $I\simeq G_1V+G_2V^2+G_3V^3$. The results of the fitting show that the magnitude of the $G_2V^2$ term is only below 1 $\%$ of $G_1$ or $G_3V^3$ in any condition of magnetic field, temperature, and barrier thickness, letting us approximately write $I\simeq G_1V+G_3V^3$.
The absence of the asymmetric term $G_2$ is reasonable considering that the MTJ film structure (CoFeB/MgO/CoFeB) is symmetric (the same material is used for both left/right electrodes), making no difference in transport characteristics for positive/negative bias.
Thus, the differential conductance at low bias regime can be well approximated by the equation below (black dashed curves in Fig. \ref{fig2}(a)).
\begin{equation}
\frac{dI}{dV}=G_1+3G_3V^2.
\label{experimental formula}
\end{equation}
Henceforth, we refer to $G_3$ as nonlinear conductance and investigate how $G_1$ and $G_3$ are correlated.

Figure \ref{fig2}(b) shows the magnetic field dependence of $G_1$ and $G_3$. The linear conductance $G_1$ takes its minimum at $\sim$0 mT and increases as the field is applied. Note that the data shown in Fig. \ref{pillar}(b) corresponds to $G_1$ via $G_1 = 1/R$. On the contrary, the magnetic field dependence of the nonlinear conductance $G_3$ takes its maximum at $\sim$0 mT and decreases as the field is increased. Unexpectedly, the behavior of $G_3$ is an upside-down reversal of that of $G_1$, suggesting the negative correlation between them.

Figure \ref{fig2}(c) shows $G_3$ as a function of $G_1$ when the magnetic field is varied from -30 mT to 30 mT.
The left- (right-) most region where $G_1$ takes its minimum (maximum) corresponds to the AP (P) magnetization configuration (see the circled area). Interestingly, a clear negative linear correlation between $G_1$ and $G_3$ is obtained for all the experimental data. The correlation is well approximated by linear function of $G_3=-kG_1+m$, where $k$ ($1/V^2$) and $m$ ($1/\Omega/V^2$) are constant coefficients.

\subsection{\label{sec:level2}Correlation plot for $G_1$ and $G_3$}
\begin{figure}[h]
\centering
\includegraphics[scale=1,pagebox=cropbox,clip]{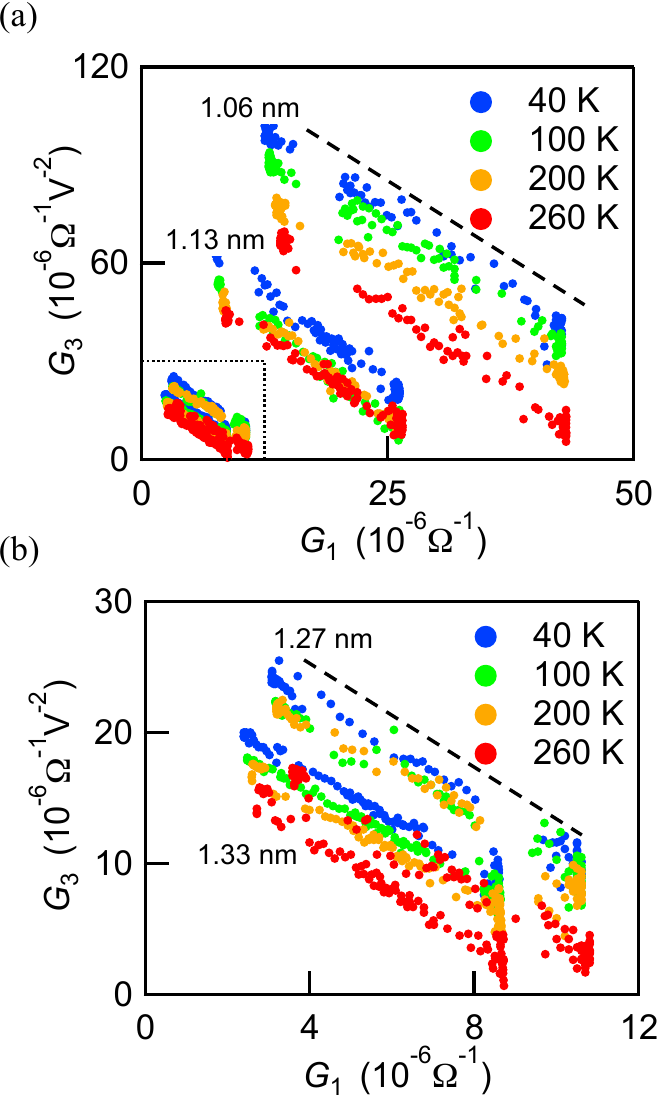}
\caption{Summary of the correlation plot for (a) $d_{\rm MgO}$ = 1.33, 1.27, 1.06 and 1.13 nm measured at $T=$ 40, 100, 200, and 260 K. The black dashed line shows the line with slope -2 ($1/\rm V^{2}$). The area inside the black dotted line is displayed in (b). (b) The correlation plot zoomed into $d_{\rm MgO}$ = 1.27 and 1.33 nm.}
\label{correlationall}
\end{figure}

Figure \ref{correlationall} shows $G_3$ vs $G_1$ correlation plots for four different thickness ($d_{\rm MgO}$ = 1.06, 1.11, 1.21, and 1.33 nm) and temperatures ($T$ = 40, 100, 200, and 260 K).
\begin{figure}[h]
\centering
\includegraphics[scale=1,pagebox=cropbox,clip]{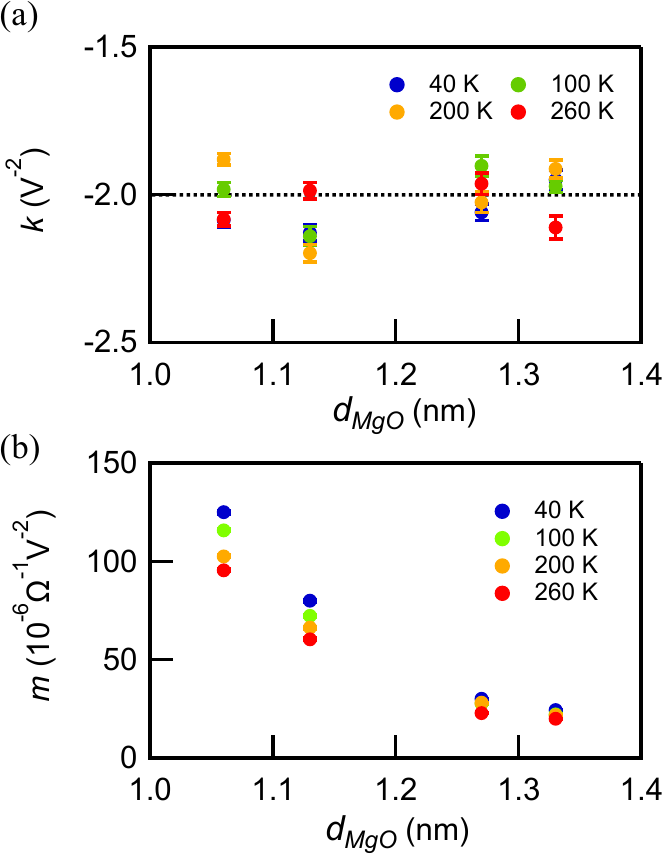}
\caption{Thickness dependence of (a) the slope $k$ and (b) the intercept $m$ extracted from the correlation plot at four different temperatures ($T=$ 40, 100, 200, and 260 K). The black dotted line in (a) shows the line with slope -2 ($1/\rm V^{2}$).}
\label{kandm}
\end{figure}
$G_1$ and $G_3$ are obtained by fitting $\frac{dI}{dV}$ between $|eV/{k_{\rm B\it}T}|\lesssim4$, while the result is not sensitive to the fitting range. As shown in Fig. \ref{correlationall}, clear negative linear correlations between $G_3$ and $G_1$ are obtained which, again, can be well approximated by $G_3=-kG_1+m$ for all the thickness and the temperature.

We perform a linear fitting of the correlation plot and summarize the results in Fig. \ref{kandm}. It turns out that the slope $k\sim2$ ($1/V^{2}$) for all the data, regardless of $T$ and $d_{\rm MgO}$ (see Fig. \ref{kandm}(a)). On the other hand, the intercept $m$ clearly depends on both $T$ and $d_{\textrm MgO}$ and becomes larger as $T$ or $d_{\textrm MgO}$ is decreased (see Fig. \ref{kandm}(b)).
Here, as the slope $k$ has a dimension of $V^{-2}$ which is independent of the geometric scale such as the junction area, it may be a universal parameter to characterize the nonlinear behavior of MTJ at low bias regime.
These results prove that the observed correlation holds in wide range of thermal fluctuation ($k_{\rm B\it}T$) and tunneling probability determined by $d_{\rm MgO}$. At the same time, it implies the relevance between the magnetization configuration and the nonlinear transport. This simple relation between the linear and the nonlinear conductance is the central experimental finding in this Article.

\section{\label{sec:level1}Discussions}
\subsection{\label{sec:level2} Barrier height modulation}
Now, we discuss the physical origin of the observed nonlinearity and the negative correlation. First of all, we examine one of the common sources of nonlinearity in tunnel junction, namely the barrier modulation effect \cite{Simmons1963,Brinkman}.

It is known that the tunneling barrier height can be modulated due to the application of the bias voltage, which makes the conductance nonlinear. One of the most accepted models is the Brinkman model \cite{Brinkman}, an extension of Simmons model \cite{Simmons1963} at low bias regime to include the barrier asymmetry. Based on the WKB approximation, the Brinkman model presents the voltage-dependent conductance $G(V)$ of a tunnel junction at low bias regime as the polynomial below.
\begin{equation}
\frac{G(V)}{G(0)}=1-\left(\frac{A_0\Delta\phi}{16\phi^{\frac{3}{2}}}\right)eV+\left(\frac{9}{128}\frac{A_0^2}{\phi}\right)(eV)^2.
\label{Brinkmaneq}
\end{equation}

In this equation, $\phi$ is the effective height (in unit of electron Volt) and $d$ is the thickness of the barrier (in unit \AA). $\Delta\phi$ is the difference of the barrier height at the interface of the left/right electrode, which is zero when the same material is used for both electrodes. $G(0)=3.16\times10^{10}\frac{\phi}{d}e^{-1.025d\phi^{\frac{1}{2}}}$ $\Omega^{-1}\textrm{m}^{-2}$ is the conductance at zero bias voltage normalized by the junction area $S$, and $A_0=\frac{4(2m_{\textrm e})^\frac{1}{2}d}{3\hbar}$
($m_{\textrm e}$ is the electron mass and $\hbar$ is the reduced Planck constant).

According to Eq. (\ref{experimental formula}) and (\ref{Brinkmaneq}), at first sight, the Brinkman model seems to give the same voltage dependence of the conductance as observed in the experiment.
Thus, we attribute the linear ($G_{1}^\textrm{B}\propto V$) and the nonlinear ($G_{3}^\textrm{B}\propto V^3$) conductance by the Brinkman model as below.
\begin{eqnarray}\label{G1cal}
  G_{1}^\textrm{B}&&=3.16\times10^{10}\frac{\phi S}{d}e^{-1.025d\phi^{\frac{1}{2}}}.  \\
  G_{3}^\textrm{B}&&=G_{1}\times\frac{9}{128}\frac{A_{0}^2S}{\phi}.
  \label{G3cal}
\end{eqnarray}
Here, the effective barrier height $\phi$ may vary with the magnetization configuration in our MTJ. This is due to the fact that the tunneling in MgO-based MTJ is dominated by different Bloch states ($\Delta_{1}$ state in P configuration and $\Delta_{5}$ state in AP configuration) and their decay rate is significantly different \cite{Butler2001,PhysRevB.63.220403}.
Now, we assume that such behaviors of the Bloch states affect the tunneling probability and modulate the effective barrier height $\phi$.
We compare $G_{1}^\textrm{B}$ and $G_{3}^\textrm{B}$ deduced from Eqs. (\ref{Brinkmaneq}), (\ref{G1cal}), and (\ref{G3cal}) with the experimental data to discuss the barrier modulation effect and the role of Bloch states in the nonlinear conductance.

\begin{figure}[h]
\centering
\includegraphics[scale=1.0,pagebox=cropbox,clip]{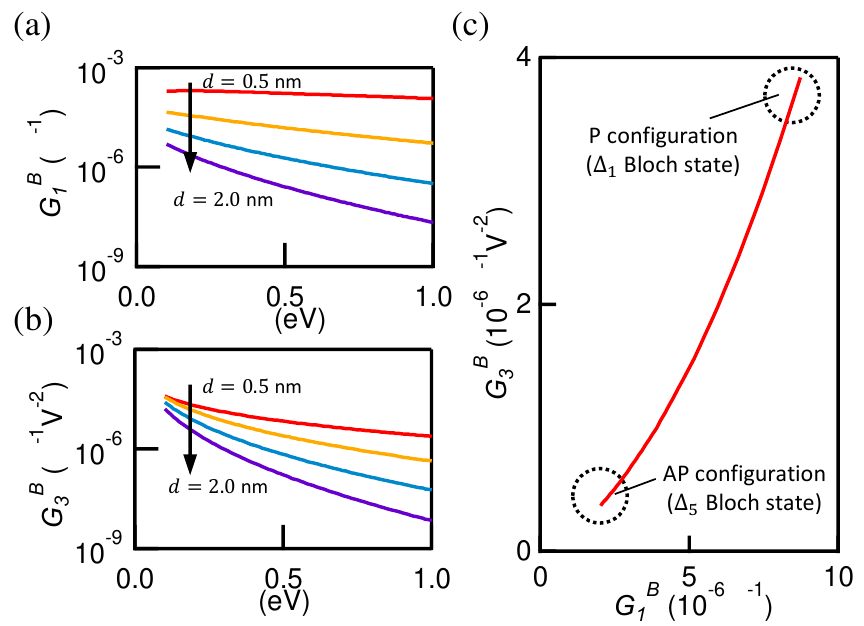}
\caption{Calculated results of (a) the linear conductance $G_{1}^\textrm{B}$ and (b) the non-linear conductance $G_{3}^\textrm{B}$ by the Brinkman model as functions of $d$ and $\phi$. (c) Correlation plot of $G_{1}^\textrm{B}$ and $G_{3}^\textrm{B}$ with changing $\phi$ from 0.3 to 1 eV ($d$ is fixed at 1.3 nm). $G_{1}^\textrm{B}$-maximum (minimum) region is marked as black dotted circle, noting P (AP) configuration and $\Delta_1$ ($\Delta_5$) Bloch state.}
\label{r1}
\end{figure}

We start with the calculation of $d$ and $\phi$ dependence of $G_1$ and $G_3$ in Eq. (\ref{G3cal}) as shown in Fig. \ref{r1}.
The barrier thickness $d$ is swept from 0.5 nm to 2 nm, and $\phi$is swept from 0.3 eV to 1 eV, both of which cover the reasonable value range for CoFeB/MgO/CoFeB.
Both $G_1$ and $G_3$ are the monotonous decreasing function of $d$ and $\phi$ (see Fig. \ref{r1}(c)).

Now, we compare these calculations with the experimental data and discuss that the Brinkman model is not sufficient to explain our experimental observation. As an example, we use the experimental data of MTJ with MgO thickness 1.33 nm as shown in Fig. \ref{fig2}.  Note that the results shown below hold true for other MgO thickness (1.06, 1.13, and 1.27 nm) in the experiment.

The nonlinearity at P configuration where the $\Delta_1$ Bloch state is dominant, the experimental value is $G_1\simeq 9\times10^{-6}$ $\Omega^{-1}V^{-1}$ and $G_3 \simeq 4\times10^{-6}$  $\Omega^{-1}V^{-2}$ respectively (see the circled area in Fig. \ref{fig2}(c) annotated ``P configuration'').
These values correspond to the ones expected for $\phi=$0.3 eV and $d$=1.3 nm in the above calculation, which give $G_1^{\rm B\it}\simeq 9\times10^{-6}$ $\Omega^{-1}V^{-1}$ and $G_3^{\rm B\it}\simeq 4\times10^{-6}$  $\Omega^{-1}V^{-2}$ (see the circled area in Fig. \ref{r1}(c) annotated “P configuration ($\Delta_1$ Bloch state)”).
Therefore, it is suggested that the barrier modulation effect by this $\Delta_1$ state is the dominant source of non-linearity at P configuration. In addition, the estimated effective barrier height and thickness ($\sim$0.3 eV and $\sim$1.3 nm) agree with those in the previous report \cite{Yuasa2004}.

Now, in the AP configuration where the $\Delta_5$ Bloch state is dominant, the experimental value is $G_1\simeq 3\times10^{-6}$ $\Omega^{-1}V^{-1}$ and $G_3\simeq 15\times10^{-6}$ $\Omega^{-1}V^{-2}$ respectively (see the circled area in Fig. \ref{fig2}(c) annotated “AP configuration”).
Here, the effective barrier height should be higher than that in P configuration because the $\Delta_5$ state decays more rapidly than $\Delta_1$ state \cite{Butler2001}.
Considering that the $G_3$ is the decreasing function of $\phi$ (see Eq. (\ref{G3cal})), however, the barrier modulation effect from this $\Delta_5$ state cannot be larger than that in the $\Delta_1$ state (e.g., $G_3\simeq 4\times10^{-6}$ $\Omega^{-1}V^{-2}$ at 20 mT in Fig. \ref{fig2}(c)).
Therefore, it is estimated that the nonlinearity due to the $\Delta_5$ state counts at most $\frac{4\times 10^{-6}}{15\times 10^{-6}}\sim 26 \%$ of the total value of $G_3$.
In addition, if we directly apply the Brinkman model to the AP configuration, we obtain $\phi=0.08$ eV and $d=2.4$ nm, which significantly deviate from the experimental condition and unrealistic.
These results suggest that Bloch states play a role in nonlinearity especially for the P configuration, while additional mechanism is required to further account for the nonlinear behavior of MTJ of whole configuration.

In addition, according to Eq.(\ref{Brinkmaneq}), both $G_{1}^\textrm{B}$ and $G_{3}^\textrm{B}$ are the monotonous decreasing functions of $d$ and $\phi$.
This means that even when $d$ and $\phi$ changes with experimental conditions (magnetic field, temperature, and MgO thickness), $G_1$ and $G_3$ change in the same way.
Therefore, as long as we consider the barrier modulation mechanism such as the Brinkman model, we would expect positive correlation between $G_1$ and $G_3$ (see Fig. \ref{r1}(c)), which is opposite to the negative one that we observed in experiment.

The model presented here does not explicitly take the effect of spin-dependent DOS into account. In principle, we can extend the Brinkman model to include such an effect. However, it turned out that our conclusion that the experimental results cannot be explained by the Brinkman model alone remains the same whether we take the effect of DOS into account or not. To keep the discussion uncomplicated, therefore, we simply show the Brinkman model without the effect of DOS here. For this reason, the estimated $\phi$ (the effective barrier height) should be different from the actual value. We note that the same conclusion can be obtained based on the Simmons model as well.

\subsection{\label{sec:level2} Control experiment}
We perform a control experiment with normal metal-insulator-ferromagnet (NIF) junction of Ta(5 nm)/Ru(10 nm)/Ta(5 nm)/Co$_{20}$Fe$_{60}$B$_{20}$ (5 nm)/MgO(1.0 nm)/Ta(5 nm)/Ru(5 nm) film. From this experiment, we can estimate the effect of impurity scattering \cite{Wei2010} because the film is prepared in the same manner as MTJ so that it has almost the same quality of tunneling barrier and impurity concentration. The sample is patterned into an elliptical pillar of 300 nm $\times$ 450 nm. Note that the area of the ellipse is twice as much as the one in the MTJ.

\begin{figure}[htbp]
\includegraphics[pagebox=cropbox,clip]{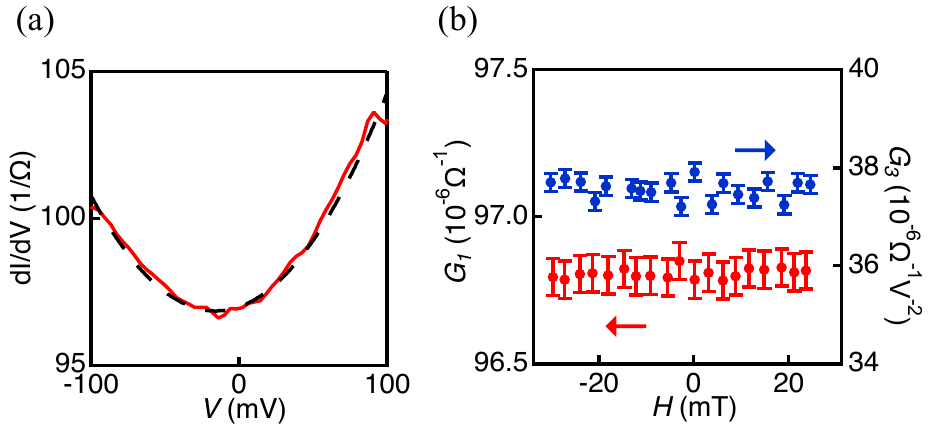}
\caption{(a) differential conductance of the NIF junction. Black dotted line is the parabolic fit. (b) Magnetic field dependence of $G_1$ and $G_3$.}
\label{control1}
\end{figure}

We measure the magnetic field dependence of the $IV$ characteristics at room temperature, which is again well fitted by a parabolic function $G=G_1+2G_2V+3G_3V^2$. Contrary to the case in MTJ, $G_{2}$ is not negligible in this result. This is consistent with the Brinkman model (see Eq. \ref{Brinkmaneq}) that expects finite $G_2$ when the materials for the left and right electrodes (CoFeB and Ta) are different.
Now we discuss the origin of the nonlinearity observed in this control experiment.
Firstly, the impurity scattering yields the conductance proportional to $|V|$ \cite {Wei2010}, which is not the case in this experiment.
Secondly, we consider the barrier modulation effect according to the Brinkman model. As shown in Fig. \ref{control1}, neither $G_1$ nor $G_3$ depends on the magnetic field in our experimental condition. Such behavior is totally different from the case with MTJ. Actually, according to Eq. (\ref{Brinkmaneq}), the estimated barrier height ($\sim$0.3 eV) and barrier thickness ($\sim$1.3 nm) are reasonable compared to the previous works \cite{Parkin2004,Yuasa2004}. This implies the nonlinear conductance in this control experiment originates from the barrier modulation effect, which does not depend on the magnetic field.


From the above discussion based on the Brinkman model and the control experiment, we can say that those mechanisms arising from the tunnel junction structure alone cannot explain the negative correlation and a certain intrinsic mechanism that stems from the spin-dependent tunneling dynamics depending on the magnetization dynamics is required.

\subsection{\label{sec:level2} Magnon-assisted tunneling}
We propose the contribution of the quasiparticles, particularly magnons to the nonlinear conductance.
It is created by the magnetization fluctuation, whose energy scale is a few tens of meV (a few hundreds of $k_\textrm{B}T$) \cite{Wei2010}.
This implies that the magnon-assisted tunneling is likely to be relevant in the observed nonlinear conductance.
In Ref. \cite{Guinea1998}, it is argued that the $G\propto V^2$ component appears when the electrons interact to magnons that exist at the junction interface. Besides, it has also been found that the contribution of such surface magnons is important in the tunnel junctions based on Co such as Co/$\rm Al_{2}O_{3}$/Co \cite{Zhang1997} and CoFeB/MgO/CoFeB \cite{Wei2010}.

\begin{figure}[h]
\begin{center}
\includegraphics[pagebox=cropbox,clip]{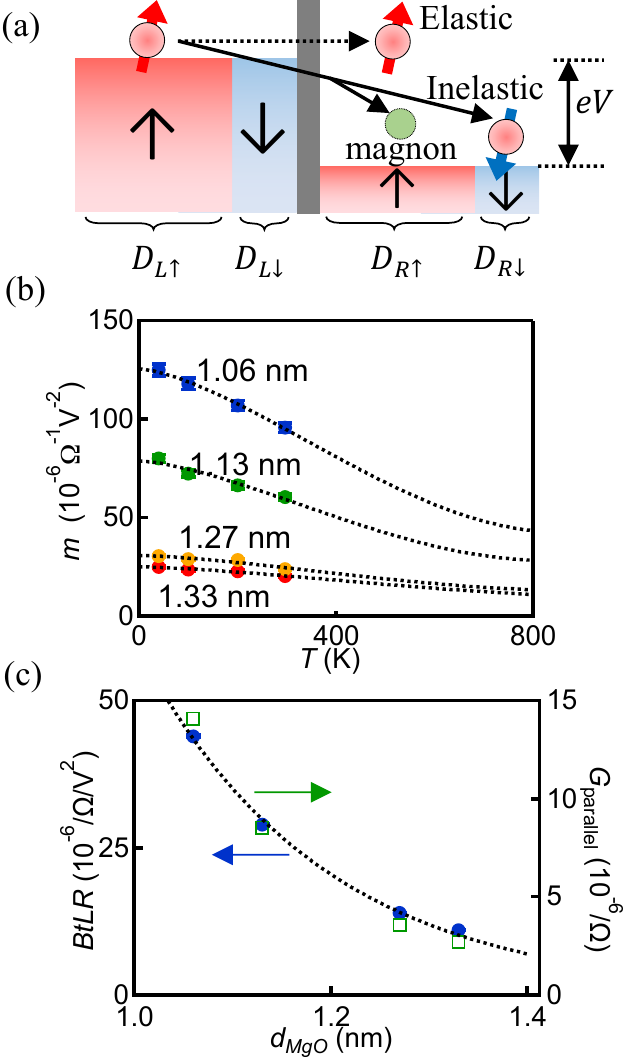}
\caption{(a) Schematic energy diagram of the MTJ. The vertical axis corresponds to the energy. Red and blue areas represent the DOS of the spin up and down electrons in each electrode respectively, separated by the insulating barrier (gray rectangle). The energy and spin conserving (``Elastic'', black dotted arrow) path and the non-conserving path (``Inelastic'', black arrows) are shown. (b) Temperature dependence of the coefficient $m$. Black dotted curves are the results of the fitting. (c) MgO thickness dependence of the coefficient $BtLR$ obtained from the fitting (left axis), plotted together with the linear conductance in P configuration ($T=260$ K) (right axis). Black broken line is the fitting with exponential function $\propto$ $e^{-d_{\rm MgO}/L}$. Here, $L\sim0.15$ nm is the constant with length dimension obtained from the fitting.}
\label{gainenzu}
\end{center}
\end{figure}

To further apply this idea to our analysis, we present a Julliére model \cite{Julliere1975} extended to include the spin-flip tunneling assisted by magnons. Fig. \ref{gainenzu}(a) illustrates the tunneling processes we consider below.
Originally, the Julliére model describes the linear current as the elastic tunneling ($\it I_{\rm elastic}$$\propto$$V$), which conserves spin and energy of electron (see the black dotted arrow annotated ``Elastic'' in Fig. \ref{gainenzu}(a)). The current is proportional to the product of the DOS with the same spin and energy,
\begin {equation}
I_\textrm{elastic}=G_1V=At(D_{L_\uparrow} D_{R_\uparrow}+D_{L\downarrow} D_{R_\downarrow})V.
\end {equation}
Here, we denote the spin-dependent density of states (DOS) at Fermi level as $D_{L_\uparrow}$, $D_{L_\downarrow}$ for the left electrode, and $D_{R_\uparrow}$, $D_{R_\downarrow}$ for the right electrode ($\uparrow$ and $\downarrow$ are spin-up and down, respectively). $A$ is a constant proportional to the junction area and $t$ is the matrix element associated with the tunneling process.

We define the tilt angle $\theta$ of the magnetization in the right electrode, seen from the left one.
$\theta$ = 0 and $\rm \pi$ correspond to P and AP, respectively.
Note that $D_{R_\uparrow}$ and $D_{R_\downarrow}$ corresponds to the magnetization direction projected to the $\theta=0$ axis, and thus is a function of $\theta$.
\begin{eqnarray}
  D_{L_\uparrow}&&=\frac{1+p}{2}L, \label{array0}\\
  D_{L_\downarrow}&&=\frac{1-p}{2}L,\\
  D_{R_\uparrow}&&=\frac{1+p}{2}R-pR\rm sin^2\frac{\theta}{2},\\
  D_{R_\downarrow}&&=\frac{1-p}{2}R+pR\rm sin^2\frac{\theta}{2}.
  \label{array1}
\end{eqnarray}
We note $D_{R_\uparrow}$ and $D_{R_\downarrow}$ at $\theta=0$ as $D^{0}_{R_\uparrow}$ and $D^{0}_{R_\downarrow}$.
The spin-polarization of left/right electrode is $p\equiv |(D_{L_\uparrow}-D_{L_\downarrow})/(D_{L_\uparrow}+D_{L_\downarrow})|=|(D^{0}_{R_\uparrow}-D^{0}_{R_\downarrow})/(D^{0}_{R_\uparrow}+D^{0}_{R_\downarrow})|$.
We also define $L \equiv D_{L_\uparrow}+D_{L_\downarrow}$
and $R\equiv D^{0}_{R_\uparrow}+D^{0}_{R_\downarrow}$.
According to the Eq. (\ref{array0}--\ref{array1}), one obtains the linear conductance as below.
\begin{eqnarray}
  G_1=&&AtLR(D_{L\uparrow}D_{R\uparrow}+D_{L\downarrow}D_{R\downarrow})\nonumber\\
  =&&AtLR\left[\frac{1-p^2}{2}+p^2\cos^{2}\frac{\theta}{2}\right].
  \label{g1}
\end{eqnarray}
This equation is a straightforward extension of the Julliére model \cite{Julliere1975} to include magnetization angle-dependence \cite{PhysRevB.39.6995}.
The energy dependence of DOS \cite{Liu2012} is neglected because the energy scale of the bias voltage ($eV\sim$ a few tens of meV) is small enough compared to the Fermi energy.

Next, we further extend the Julliére model to describe the nonlinear conductance ($G_3$). We assume that the nonlinear current corresponds to the inelastic tunneling current ($I_{\rm inelastic}$$\propto$$V^3$), where the electron changes its energy and spin (see the black arrow in Fig. \ref{gainenzu}(a) annotated ``Inelastic'').
We describe the nonlinear conductance using the products of DOS with opposite spins,
\begin {equation}
I_{\rm inelastic}=G_3 V^3=BtLR(D_{L_\uparrow} D^*_{R_\downarrow}+D_{L_\downarrow} D^*_{R_\uparrow})\it V^{\rm 3}.
\end {equation}
Here, we set an ad-hoc assumption on DOS:
\begin{eqnarray}
  D^*_{R\uparrow}\equiv&& D_{R\uparrow}-D^{0}_{R\uparrow}=-pR\sin^{2}\frac{\theta}{2}.\\
  D^*_{R\downarrow}\equiv&& D_{R\downarrow}-D^{0}_{R\downarrow}=pR\sin^{2}\frac{\theta}{2}.
\end{eqnarray}
In this assumption, the DOS in the P configurations is subtracted from the original one, which enables us to subtract the nonlinearity that is already present at P configuration. This allows us to focus on the continuous changes of nonlinearity between P and AP configurations. We will examine the relevance of this assumption later.
The inelastic term is simplified to,
\begin{eqnarray}
  G_3=&&BtLR\left(D_{L_\uparrow} D^*_{R_\downarrow}+D_{L_\downarrow} D^*_{R_\uparrow}\right)\nonumber\\
  =&&BtLRp^2\sin^2\frac{\theta}{2}.
  \label{g3}
\end{eqnarray}
In this expression, $G_3$ becomes minimum in P configuration ($\theta=0$) in agreement with the observation shown in Fig. \ref{fig2}.

From Eq. (\ref{g1}) and (\ref{g3}) the above expressions of $G_1$ and $G_3$, we obtain,
\begin {equation}
G_3=-\frac{B}{A}G_1+\frac{BtLR}{2}(1+p^2)=-kG_1+m.
\end {equation}

Here,
\begin{eqnarray}
k=&&B/A.\\
m=&&\frac{BtLR}{2}(1+p^2).
\label{m}
\end{eqnarray}
According to this formula, only the intercept $m$ depends on the barrier thickness and the temperature, while the slope $k$ is a constant that is independent to them.
This indicates the linear negative correlation between $G_1$ and $G_3$, which qualitatively explains the experimental observation in Fig. \ref{correlationall}.

Now, let us further try to apply the above model to the experimental data. According to this model, only the intercept $m$ depends on temperature through polarization $p$.
Most simply, $p$ is assumed to obey the Bloch's law \cite{Bloch1930,Shang1998}
\begin{equation}
  p=p_0\left[1-\left(\frac{T}{T_c}\right)^\frac{3}{2}\right].
  \label{p}
\end{equation}
This describes the change of magnetization due to the magnon excitation ($p_0$ is the spin polarization at 0 K and $T_c$ is the Curie temperature).
The details of the magnon in this system are included in these macroscopic parameters.
Note that such temperature dependence is introduced through the definition of the nonlinear conductance such as seen in Eq. (\ref{g3}).
Conversely, if we naively defined  $G_3=Bt(D_{L_\uparrow} D_{R_\downarrow}+D_{L_\downarrow} D_{R_\uparrow})$, the correlation would become $G_3=-\frac{B}{A}G_1+\frac{BLRt}{2}$ and no temperature dependence would appear in the intercept, which fails to explain the experimental observation. This is the reason why we adopted the ad-hoc assumption to interpret the behavior of $m$.

Thus, according to Eq. (\ref{m}) and Eq. (\ref{p}), the intercept $m$ is obtained as below.
\begin{equation}
m=\frac{BtLR}{2}\left[1+p_0^{2}\left(1-\left(\frac{T}{T_c}\right)^{\frac{3}{2}}\right)^2\right].
\label{mnoformula}
\end{equation}
We show the results of the fitting of $m$ with Eq. \ref{mnoformula} in Fig. \ref{gainenzu}(b). The parameters ($B, t, L, R, p_0$, and $T_c$) are set as free parameters. The estimated Curie temperature $T_c\sim$800 K and spin polarization $p_0\sim0.7$ are reasonable for the CoFeB thin film \cite{doi:10.1063/1.4985720,Satoh2018,Teixeira2010}.
The agreement between the data and the fitting is not sensitive to the slight change of $T_c$ and $p_0$.
Furthermore, as shown in Fig. \ref{gainenzu}(c), the coefficient $BtLR$ exponentially decreases with increasing $d_{\rm MgO}$.
This behavior is most likely attributed to the $d_{\rm MgO}$ dependence of the tunneling probability $t$.
In fact, as shown in Fig. \ref{gainenzu}(c), the linear conductance $G_1$$\propto$$t$ decreases with the same exponential function with $BtLR$. This agreement ensures that the $t$ obtained from the model actually corresponds to the transmission probability.

The above model is a simple one but the obtained relation consistency between $G_1$ and $G_3$ seems to suggest that this captures the essential characteristics of the experimental results. The model is nevertheless still phenomenological that it cannot explain, for example, the microscopic origin of the constant $k$ and the validity of the ad-hoc assumption. For further study, the calculation based on the microscopic Hamiltonian of electron tunneling and the electron-magnon interactions (e.g. non-equilibrium Green function) would be a powerful approach.

\section{\label{sec:level1} Conclusion}
In conclusion, we have investigated the electron transport right beyond linear response regime in MTJ at low bias regime. We have found a clear negative correlation between the linearity and the nonlinearity. This cannot be fully attributed to the barrier modulation effect (Brinkman model), the rest of which we attribute to the magnon-assisted tunneling through our phenomenological model by extending Julliére model. Our findings contribute to the deeper understanding of the MTJ based on the rigorous understanding of the linear response regime.

\begin{acknowledgements}
  This work is partially supported by JSPS KAKENHI Grant Nos. JP17K18892, JP18J20527, JP19H00656, JP19H05826, JP16H05964, and JP26103002. The authors acknowledge T. Kato, Y. Ominato, and M. Matsuo for fruitful discussions. The authors also acknowledge M. Takahagi for technical support.
\end{acknowledgements}
\bibliography{MTJ_cite}

\end{document}